\documentclass{aa}  

\usepackage{graphicx,amsmath}
\usepackage{txfonts}
\usepackage{color}
\usepackage{multirow}

\usepackage{tikz}
\usepackage{comment}
\usepackage{hyperref}
\usepackage{placeins}

\newcommand{\ve}[1]{ {\mathbf{#1}} }
\newcommand{\bmu}{\boldsymbol{\mu}}
\newcommand{\balpha}{\boldsymbol{\alpha}}
\def\W{\ve{W}}
\def\H{\ve{H}}

\usepackage{ulem}
\usepackage{booktabs}

\begin{document} 
\sloppy

\title{PDRs4All}
\subtitle{XX. \textit{Haute Couture}: Spectral stitching of JWST MIRI-IFU cubes \\with matrix completion}

   \author{Amélie Canin \inst{1} \and
            Cédric Févotte \inst{2}  \and
            Nicolas Dobigeon \inst{2}\and 
            Dries Van De Putte \inst{3} \and
            Takashi Onaka \inst{4} \and
            Olivier Berné \inst{1}}

 \institute{Institut de Recherche en Astrophysique et Plan\'etologie (IRAP), CNRS, CNES, Université de Toulouse, France \and Institut de Recherche en Informatique de Toulouse (IRIT), CNRS, Toulouse INP, Université de Toulouse, France
 \and University of Western Ontario, Canada \and Department of Astronomy, Graduate School of Science, The University of Tokyo, Japan}
 

   \date{}

 
  \abstract
{MIRI is the imager and spectrograph covering wavelengths from $4.9$ to $27.9$ $\mu$m {onboard} the James Webb Space Telescope (JWST). The Medium-Resolution Spectrometer (MRS) consists of four integral field units (IFU), each of which has three sub-channels. The twelve resulting spectral data cubes have different fields of view, spatial, and spectral resolutions. The wavelength range of each cube partially overlaps with the neighboring bands, and the overlap regions typically show flux mismatches which have to be corrected by spectral stitching methods. \textit{Stitching} methods aim to produce a single data cube incorporating the data of the individual sub-channels, which requires matching the spatial resolution and the flux discrepancies. We present \textit{Haute Couture}, a novel stitching algorithm which uses  \textit{non-negative matrix factorization} (NMF) to perform a \textit{matrix completion}, where the available MRS data cubes are treated as twelve sub-matrices of a larger incomplete matrix. Prior to matrix completion, we also introduce a novel pre-processing to homogenize the global intensities of the twelve cubes. Our pre-processing consists {in} jointly optimizing a set of global scale parameters that {maximize} the fit between the cubes where spectral overlap occurs. We apply our novel stitching method to JWST data obtained as part of the PDRs4All observing program of the Orion Bar, and produce a uniform cube reconstructed with the best spatial resolution over the full range of wavelengths.}

    
   \maketitle


\section{Introduction} \label{sec:intro}

\begin{figure}[t]
    \centering
    \includegraphics[width=1\linewidth]{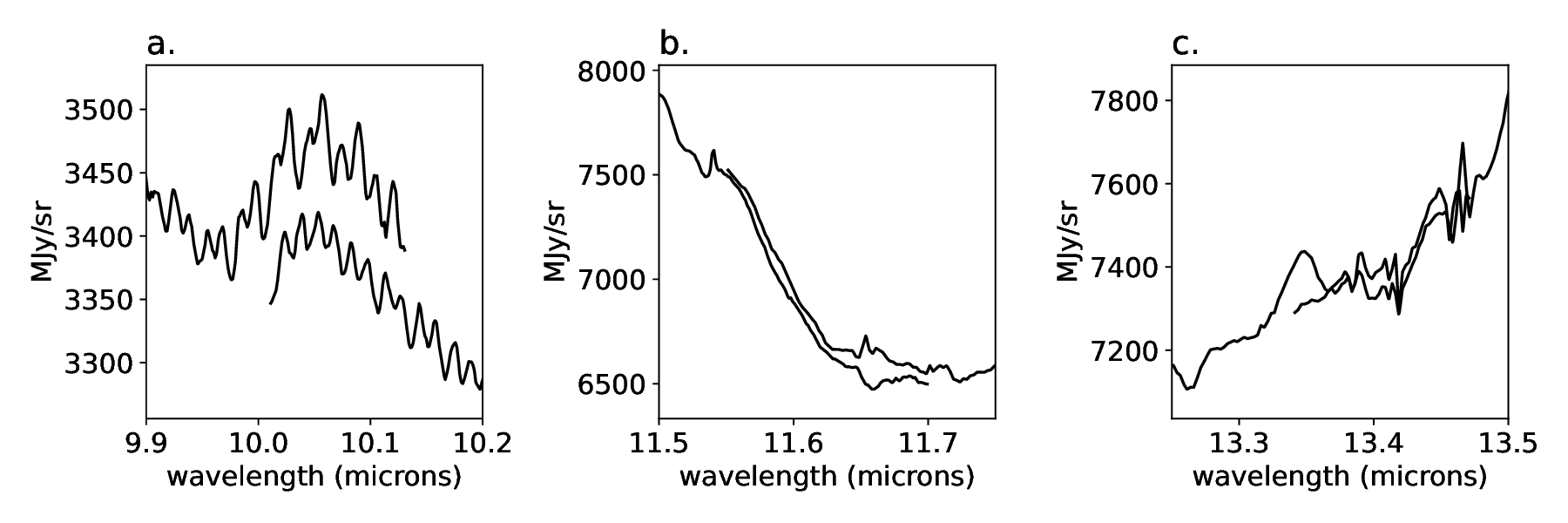}
    \caption{Examples of intensity gaps observed in overlapping areas of adjacent cubes, obtained from MIRI-MRS data collected within the PDRs4All program (see Section \ref{sec:data_preprocessing} for more details). The spectra displayed correspond to the average spectra over the entire field of view for two overlapping sub-channels. Panel a: channel 2-\textit{medium} and 2-\textit{long}. Panel b: channel 2-\textit{long} and 3-\textit{short}. Panel c: channel 3-\textit{short} and 3-\textit{medium}.}  
    \label{fig:gaps}
\end{figure}

The James Webb Space Telescope (JWST, \citet{gardner2006}) is an infrared telescope launched in 2021.
Four instruments are onboard: the Mid-Infrared Instrument (MIRI, \citealt{rieke2015,wright2023}), the Near Infrared Camera (NIRCam), the Near Infrared Spectrograph (NIRSpec), and the Near Infrared Imager and Slitless Spectrograph (NIRISS). MIRI provides images and spectroscopic data through different observing modes covering $4.9$ to $27.9$ $\mu$m. 
The Medium-Resolution Spectroscopy mode of MIRI (MRS, \citealt{wells2015,argyriou2023}) uses four Integral Field Units (IFUs), each one {covering a portion of the wavelength range referred to as channel}. Each {of these channels} is subdivided {into} three sub-bands (\textit{short}, \textit{medium} and \textit{long}). {Consecutive channels and bands share some information thanks to spectral overlaps. Besides, along the channels,} the field of view increases  {whereas} the spatial resolution decreases. The Science Calibration Pipeline provided by the Space Telescope Science Institute thus produces {a} total of twelve data cubes {with different spatial coverages and resolutions}. {Another challenge raised by the MIRI data acquisition process results from intensity gaps which affect the spectral measurements between adjacent cubes. These intensity gaps, visible from typical spectra reproduced in Figure \ref{fig:gaps}, may be due to calibration mismatches}. {To make MIRI data easier to use, a convenient data product would be a singular data cube free of spectral intensity gaps, spanning the full spectral range, covering a common field of view, which preserves the highest spatial resolution attainable in each band.}

\begin{figure}[t]
    \centering
    \includegraphics[width=1\linewidth]{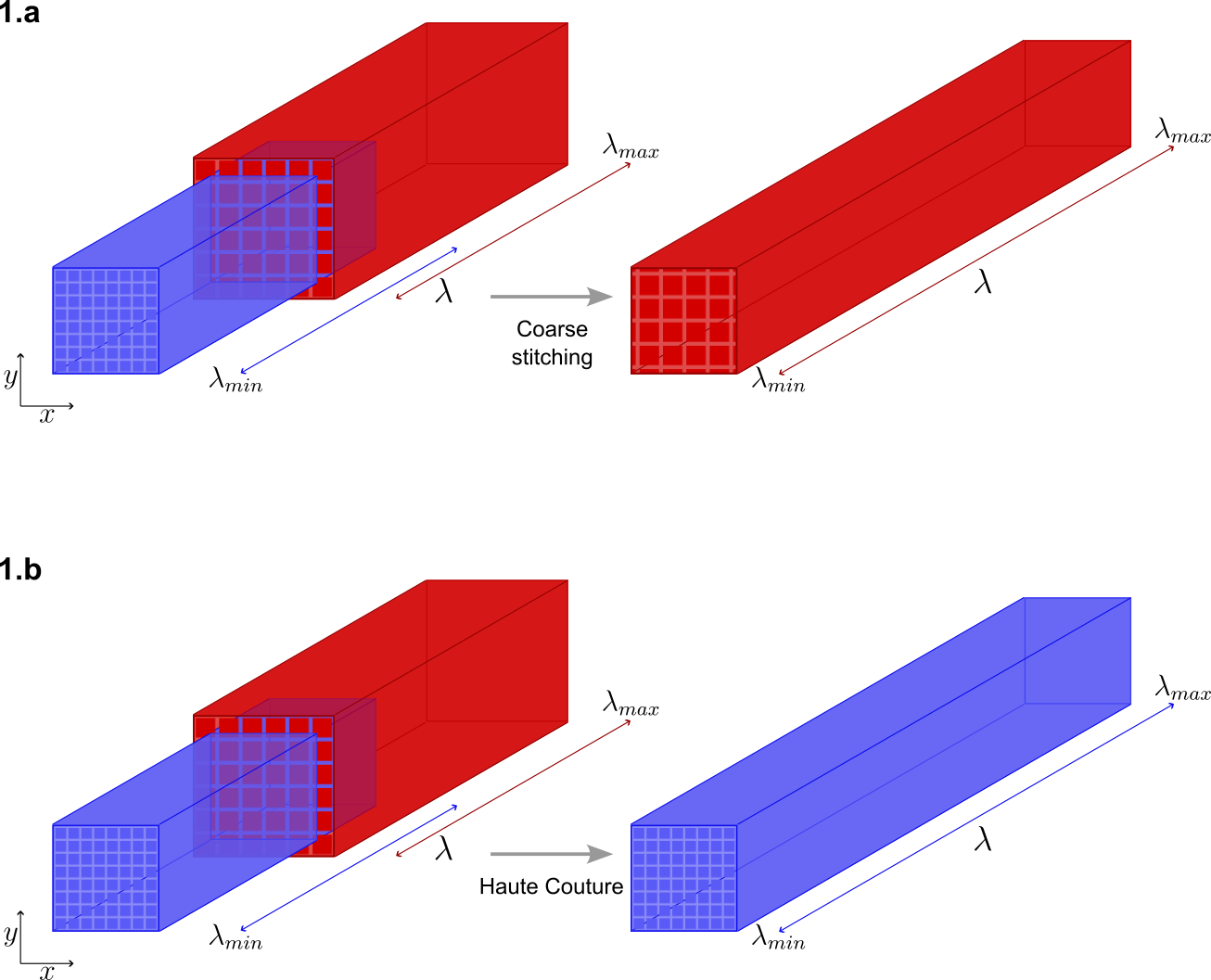}
    \caption{Illustration of the stitching problem for two cubes. On the left, we observe that the shortest-wavelength cube in blue has a better spatial resolution (as illustrated by the finer grid) and smaller field of view than the longest-wavelength cube in red. The space $\times$ wavelength supports of the cubes partially overlap. The coarse stitching procedure shown in Figure 1.a sacrifices spatial resolution, while \textit{Haute Couture} shown in Figure 1.b enables the reconstruction of a cube with best spatial resolution over the full range of wavelengths.}
        \label{fig:concept}
\end{figure}

{The process of assembling individual spectral data cubes is often referred to as {\it stitching}. Because of the aforementioned challenges resulting from the unconventional acquisition implemented by MIRI, stitching raises the {following questions \textit{i)} how to assemble intensity gap-free spectra from the twelve data cubes? and \textit{ii)} how to deal with the distinct spatial sampling (which varies with 1/$\lambda$)}? A commonly employed stitching method  consists in reprojecting all the spectral data onto the poorest resolution spatial grid. This was for instance applied on data from the Infrared Spectrograph 
\citep{houck2004infrared} taken during the Spitzer mission \citep{werner2004spitzer}.
This was done by first convolving all the individual cubes with the point spread function (PSF) of the largest wavelength (corresponding to the lowest resolution) in order to reach the same spatial resolution over the full spectral range (see, e.g., \citealt{berne2007,sandstrom2010}). This procedure, referred  to as ``coarse stitching'' hereafter, is illustrated in Figure \ref{fig:concept}.a when considering  two data cubes. Unfortunately, this strategy leads to a regrettable loss of valuable spatial information.}

To overcome the limitations inherent to { coarse stitching} strategy, this paper presents a novel method, coined as \textit{Haute Couture}, that enables  stitching while preserving spatial resolution. The proposed method performs the reconstruction of a full data cube at the best spatial resolution over the full spectral range, as illustrated in Figure \ref{fig:concept}.b. We first show that the data recorded in the twelve sub-channels can be rearranged as twelve sub-matrices of a larger matrix with missing values. Interestingly, recovering these missing values amounts to reconstructing a data cube at the finest spatial resolution over the full spectral range.
\textit{Haute Couture} exploits the overlapping spectral information  between adjacent channels and bands to frame the stitching task as a {\it matrix completion} problem, which can then be solved by {\it nonnegative matrix factorization} (NMF). Besides, to avoid the issues caused by intensity gaps between consecutive channels and bands, prior to matrix completion,  \textit{Haute Couture} applies a novel pre-processing to homogenize the global intensities along the twelve cubes. This pre-processing consists of jointly optimizing a set of global scaling parameters to maximize the fit between the cubes where spectral overlap occurs. It is worth noting that this pre-processing is an independent procedure that can be combined with any stitching method.

Following the presentation of methodological compounds of \textit{Haute Couture} in Section~\ref{sec:method},  MIRI-MRS data observed as part of the PDRs4All program on the Orion Bar are presented in Section~\ref{sec:result} and stitched in Section \ref{sec:results2}. In particular, we demonstrate the ability of \textit{Haute Couture} to reconstruct a uniform cube with the best spatial resolution over the full range of wavelengths.

\begin{table}[t]
    \begin{center}
        \begin{tabular}{cccc}
            \toprule
             & Sub-channel  & $(c,b)$ & $\lambda$-range ($\mu$m)  \\ \hline
            \multirow{3}{*}{channel 1} & \textit{short} & $(1,\texttt{s})$ & 4.90 - 5.74 \\
                & \textit{medium} & $(1,\texttt{m})$ & 5.66 - 6.63 \\
                & \textit{long} & $(1,\texttt{l})$ & 6.53 - 7.65 \\ \hline
            \multirow{3}{*}{channel 2} & \textit{short} & $(2,\texttt{s})$ & 7.51 - 8.77 \\
                & \textit{medium} & $(2,\texttt{m})$ & 8.67 - 10.13 \\
                & \textit{long} & $(2,\texttt{l})$  & 10.01 - 11.70 \\ \hline
            \multirow{3}{*}{channel 3} & \textit{short} & $(3,\texttt{s})$  & 11.55 - 13.47 \\
                & \textit{medium} & $(3,\texttt{m})$ & 13.34 - 15.57 \\
                & \textit{long} & $(3,\texttt{l})$ & 15.41 - 17.98 \\ \hline
            \multirow{3}{*}{channel 4} & \textit{short} & $(4,\texttt{s})$  & 17.70 - 20.95 \\
                & \textit{medium} & $(4,\texttt{m})$ & 20.69 - 24.48 \\
                & \textit{long} & $(4,\texttt{l})$ & 14.40 - 27.90 \\ 
                \bottomrule
        \end{tabular}
    \end{center}
    \caption{Name and wavelength range of the MIRI-MRS channels and sub-channels.}
    \label{tab:ch_table}
\end{table}

\section{ \textit{Haute Couture}} \label{sec:method}

\subsection{Spectral stitching as a matrix completion problem} \label{sec:completion}

As introduced in Section~\ref{sec:intro}, MIRI-MRS acquires spectra in four channels indexed by $c$ ($c=1,\ldots,4$). Each channel is divided into three sub-channels, referred to as \textit{short}, \textit{medium} and \textit{long} and shortened as $\texttt{s}$, $\texttt{m}$ and $\texttt{l}$, respectively. The spectral ranges of each sub-channel are reported in Table \ref{tab:ch_table}. Each sub-channel produces a space $\times$ wavelength cube as schematized in Figure~\ref{fig:concept} where two of these cubes are depicted in blue and red. In the remainder of the paper, the spatial organization of the data is ignored and we unfold each cube with respect to the spatial dimension (i.e., pixels). Thus the data associated with each cube can be represented as a matrix indexed by pixels and wavelengths. Formally, the matrix which gathers the data collected in sub-channel $b \in \left \{ \texttt{s}, \texttt{m}, \texttt{l} \right \} $  of channel $c$ ($c=1,\ldots,4$) is denoted as $\mathbf{X}^{c,b} \in \mathbb{R}_+^{\Lambda^{c,b} \times P^{c,b}}$ where its rows index wavelengths while its columns index pixels. Specifically, the matrix $\mathbf{X}^{c,b}$ associated with the sub-channel $(c,b)$ gathers spectra composed of $\Lambda^{c,b}$ wavelengths and acquired over $P^{c,b}$ spatial pixels.

\begin{figure}[t]
    \centering
    \includegraphics[width=.95\columnwidth]{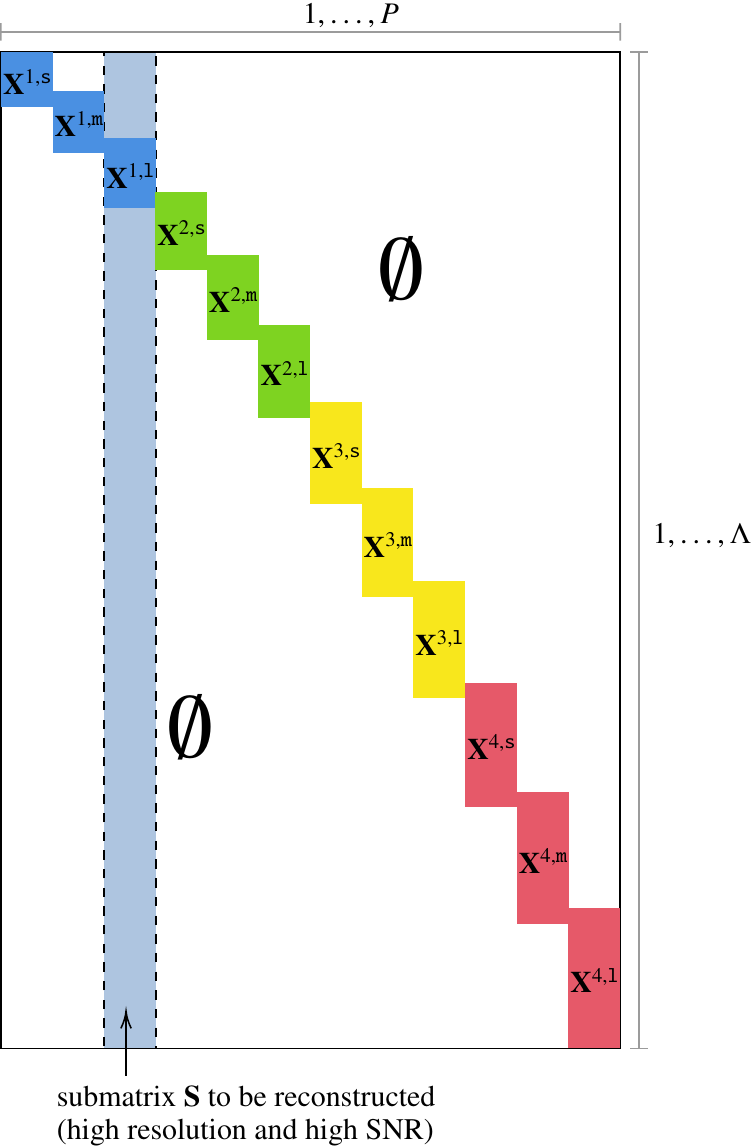}
        \caption{The twelve unfolded cubes provided by the twelve sub-channels can be organized as sub-matrices of a larger matrix $\mathbf{X}$ with {an almost} block-diagonal structure. The wavelength range of consecutive blocks overlap for a few rows, corresponding to spectral bands observed by consecutive sub-channels. Blue blocks represent channel 1, green ones channel 2, yellow ones channel 3 and red ones channel 4. White and light blue parts correspond to unobserved coefficients. The {light blue area} corresponds to the area we are specifically interested in reconstructing.}
    \label{fig:matrice}
\end{figure}

The resulting twelve sub-matrices can be arranged according to a suitable manner to form a larger matrix $\ve{X} \in \mathbb{R}_+^{\Lambda \times P}$, as represented in Figure~\ref{fig:matrice}, where $P = \sum_{c=1}^4\sum_{b\in \left \{ \texttt{s}, \texttt{m}, \texttt{l} \right \}} P^{c,b}$ and $\Lambda$ is the total number of wavelengths over the spectral range covered by MIRI-MRS. It is worth noting that the spectral range of adjacent sub-matrices overlap for a few wavelengths, i.e., $\Lambda \neq \sum_{c=1}^4\sum_{b\in \left \{ \texttt{s}, \texttt{m}, \texttt{l} \right \}} \Lambda^{c,b}$. This results in an almost block diagonal structure of the matrix $\ve{X}$. In this figure, the white parts of the matrix correspond to unavailable (i.e., unobserved) data, specified by the symbol $\emptyset$. Because of these missing coefficients, the matrix $\ve{X}$ is said to be {\it incomplete}.

The key rationale of the proposed stitching approach lies on the following insight: recovering all the missing coefficients of this matrix $\ve{X}$ would provide a comprehensive nay redundant spatial-spectral description of the scene observed by MIRI-MRS. In particular, recovering the values of the submatrix $\ve{S}$ corresponding to the blue shaded area in Figure~\ref{fig:matrice} would produce a data cube at the spatial resolution of the channel 1-{\it long} (i.e., $c=1$ and $b=\texttt{l}$) over the full spectral extent offered by MIRI-MRS, which is the main objective of this work. In other words, the spectral stitching task can be formulated as a {\it matrix completion} problem, i.e., recovering the missing data in the matrix $\ve{X}$. In this work, we target the channel 1-{\it long} to define the spatial and spectral resolutions of the stitched data because this leads to the best trade-off between high spatial resolution and high signal-to-noise ratio (SNR). However this arbitrary choice can be lifted since the proposed method will complete the full matrix $\mathbf{X}$.

The following paragraphs details the proposed two-step procedure to stitch the MIRI-MRS data while accommodating  for intensity gaps between adjacent data cubes. This procedure consists in first rescaling the individual sub-matrices $\mathbf{X}^{c,b}$ homogeneously to correct the intensity gaps (see Section \ref{sec:scaling}) and then filling  the  matrix $\ve{X}$ composed of the resulting intensity-corrected sub-matrices (see Section \ref{sec:mcomp}).

\subsection{Intensity gaps correction} \label{sec:scaling}

To mitigate the intensity gaps highlighted in Figure~\ref{fig:gaps}, we introduce a novel pre-processing to homogenize the global intensities of the twelve cubes. This pre-processing consists in jointly adjusting a set of global scale parameters to maximize the fit between the parts of the cubes where spectral overlaps occur. Let us denote by $\alpha^{c,b}$ the scale parameter to be applied to the entries of the  matrix $\mathbf{X}^{c,b}$. As an example, referring to Figure~\ref{fig:matrice}, the scale parameter, say, $\alpha^{2,{\tt l}}$, is computed by maximizing the fit between, on the one hand, the first rows of $\ve{X}^{2,{\tt l}}$ and the last rows of $\ve{X}^{2,{\tt m}}$ and, in the other hand, the last rows of $\ve{X}^{2,{\tt l}}$ and the first rows of $\ve{X}^{3,{\tt s}}$.  Because the matrices $\ve{X}^{2,{\tt m}}$ and $\ve{X}^{3,{\tt s}}$ should be themselves rescaled by the unknown scale parameters $\alpha^{2,{\tt m}}$ and $\alpha^{3,{\tt s}}$, the optimal values of the whole set of scale parameters are inter-dependent and they have to be adjusted jointly. Thankfully, when the fit is measured by the squared Euclidean distance between the portions of shared spectra, this optimization problem has a closed-form solution. 

In practice at least one sub-channels is chosen as a reference to make the problem well-posed (otherwise, a trivial but inappropriate solution would be setting $\alpha^{c,b} = 0$ for all channels and sub-channels). In the experimental results conducted in Section \ref{sec:result}, we fix $\alpha^{1,{\tt s}} = \alpha^{4,{\tt l}} =1$, and maximize the fit with respect to  the ten remaining scale parameters. This amounts to trusting the calibration of the data observed in the lowest and highest ends of the wavelength range, and update the scales of the cubes in between. Note however that any other choice is possible. Technical details and the expressions of the resulting optimal scale parameters are reported in Appendix~\ref{appx:opti}.

\subsection{Matrix completion} \label{sec:mcomp}

As explained in Section~\ref{sec:completion}, recovering the missing coefficients in the full matrix $\ve{X}$ displayed in Figure~\ref{fig:matrice} can be interpreted as a {\it matrix completion} problem. Matrix completion aims at predicting missing values given observed values and the assumption of a latent matrix structure (see, e.g., \cite{Chi2019}). In our case, and as in many other settings, it makes sense to assume that $\ve{X}$ has a {\it low-rank} structure, an assumption which has already underpinned several techniques of infrared spectroscopic data processing such as source separation \citep{rapacioli2005spectroscopy, berne2007} or data fusion \citep{berne2010non, guilloteau2020hyperspectral, guilloteau2020simulated}. It consists in imposing that the columns of $\ve{X}$ can be explained by linear combinations of elementary spectra embedded in noise. To further exploit the non-negativity of the observed spectra, we propose to perform this matrix completion by resorting to the technique of non-negative matrix factorization (NMF) \citep{lee99,Smaragdis2014}. In other words, the {\it missing entries} are approximated by $x_{ij} \approx [\ve{W} \ve{H}]_{ij}$ where $\ve{W}$ and $\ve{H}$ are non-negative matrices of size $\Lambda \times K$ and $K \times P$ estimated from the low-rank approximation of the {\it observed entries}. More precisely, denote by ${\cal O}$ the set of row and column indices of the observed values in $\ve{X}$, i.e., the non-white part of $\ve{X}$ in Figure \ref{fig:matrice} (or equivalently the supports of the sub-matrices $\ve{X}^{c,b}$). We want to estimate the two matrices $\ve{W}$ and $\ve{H}$ by solving the following optimization problem
\begin{equation} \label{eq:nmf}
    \min_{\W,\H \ge 0} \sum_{(i,j) \in {\cal O}} d(x_{ij}| [\W \H]_{ij})
\end{equation}
where $d(u|v)$ denotes a discrepancy measure between the non-negative numbers $u$ and $v$. In this study we leverage the work of \cite{fevotte2011} that proposed easy-to-implement and efficient multiplicative updates for estimating $\W$ and $\H$ when $d(\cdot|\cdot)$ is chosen as a so-called $\beta$-divergence. The $\beta$-divergence is a continuous family of measures of fit governed by a single shape parameter $\beta \in \mathbb{R}$, that takes well-known divergences as special cases, namely the generalized Kullback-Leibler and Itakura-Saito divergences ($\beta =1$, $\beta=0$, respectively) and the square Euclidean distance ($\beta=2$). Given a couple of matrices $\hat{\W}$ and $\hat{\H}$ that solve the minimization problem \eqref{eq:nmf}, the missing coefficients of $\ve{X}$ can be reconstructed as $\hat{x}_{ij} = [\hat{\W} \hat{\H}]_{ij}$ for $(i,j) \not\in {\cal O}$.

Like most NMF techniques, the algorithm tailored by \cite{fevotte2011} is a descent algorithm that relies on alternating updates of $\W$ and $\H$. Because of the bilinearity induced by the product $\W \H$, the objective function to minimize in Equation~\eqref{eq:nmf} is non-convex and the algorithm is likely to produce local solutions that depend on the chosen starting point. As such, the initialization of the descent algorithm is an issue and we will investigate several strategies in the experimental section.


\section{Application to MIRI-MRS data of PDRs4All ERS program} \label{sec:result}
\subsection{Data and preprocessing} \label{sec:data_preprocessing}

\paragraph{Data.} 

\begin{figure}
    \centering
    \includegraphics[width=1\linewidth]{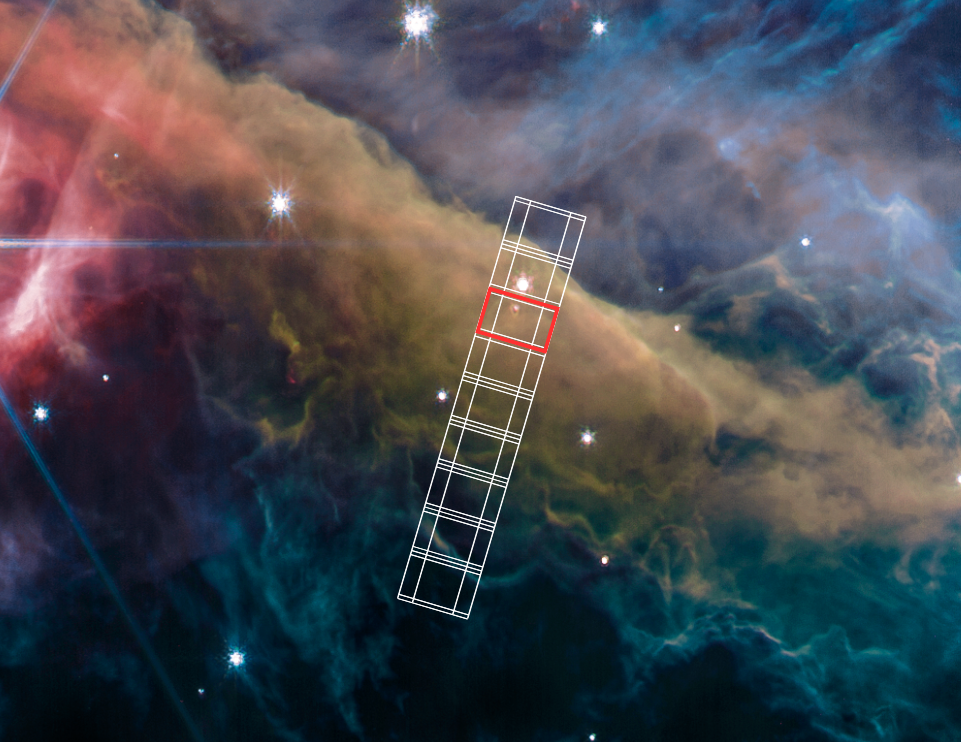}
    \caption{Colorized image NIRCam of the Orion Bar of the PDRs4All ERS program (\cite{berne2024}). Filters F140M and F210M are in blue; F277W, F300M, F323N and F335M in green; F405N in orange; and F444W, F480M and F470N in red. The pattern in white is the MIRI-MRS mosaic footprint. The red box corresponds to the pointing used in this article.}
    \label{fig:nircam}
\end{figure}

\begin{figure*}
    \centering
    \includegraphics[width=1\linewidth]{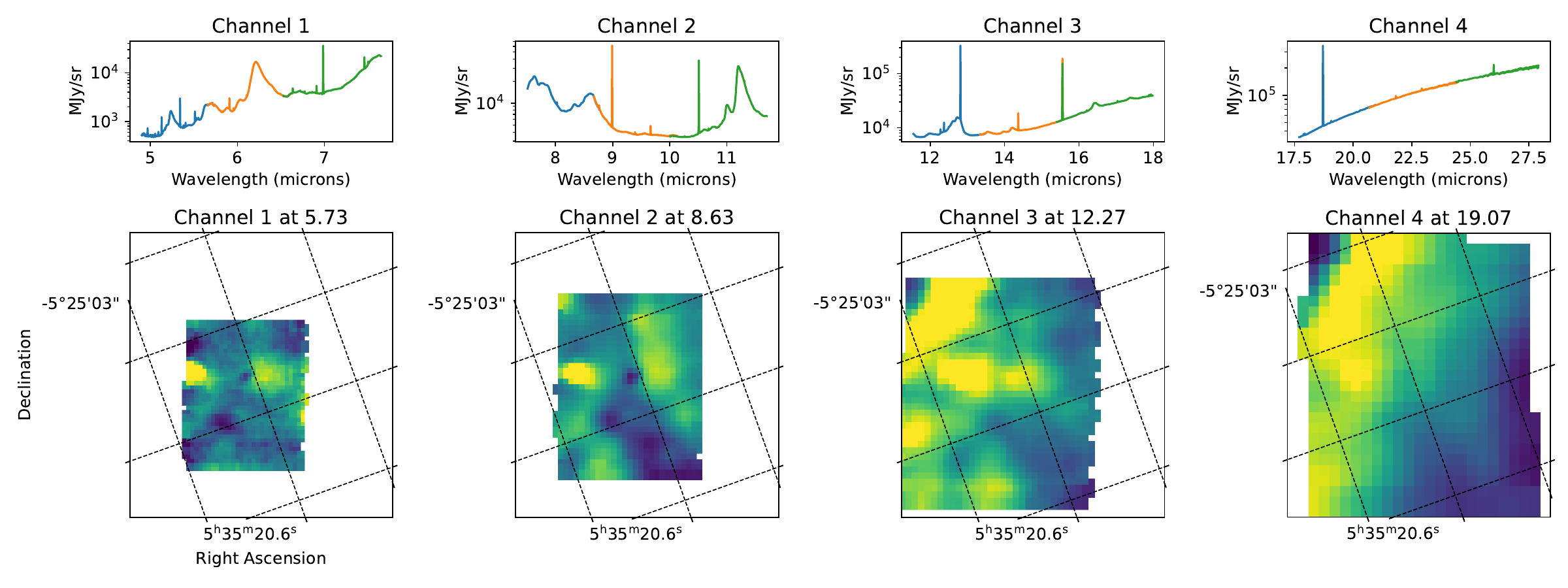}
    \caption{The first row presents the average spectra over the entire field of view of each channel (shown in the second row), with a different color (blue, orange, green) for each sub-channel (\textit{short}, \textit{medium}, \textit{long}). The second row presents the MIRI-MRS images at different wavelengths  (the images are aligned). It can be observed that the field of view increases with the channel, but that the spatial resolution deteriorates.}
    \label{fig:data}
\end{figure*}

The observations used in this article are part of the Early Release Science (ERS) program PDRs4All \citep{berne2022} obtained in January 2023. This program observed the Orion Bar with MIRI-MRS in a $1\times9$ pointing mosaic in the four channels. The mosaic is depicted in white in Figure \ref{fig:nircam}. Data was acquired with 47 groups per integration, 1 integration and 4 dithers using the FASTR1 readout. Data was presented by \citet{chown2024} and \citet{putte2024}. The experiments reported in this paper have used a single mosaic tile containing the protoplanetary disk d203-506 (highlighted in red in Figure \ref{fig:nircam}) which was also previously studied by \citet{berne2023} and \citet{zannese2024}. Figure \ref{fig:data} presents the MIRI-MRS original data cubes with the initial field of view in each channel.

\paragraph{Data reduction pipeline.} To illustrate the relevance of the proposed stitching method on a set of data representative of the current quality of JWST data products, the MIRI MRS data was reprocessed starting from the `\texttt{uncal}' files downloaded from MAST.
For this purpose we used the scripts provided by the PDRs4All data reduction team and available from a public repository\footnote{\url{https://github.com/PDRs4All/PDRs4All}}, in combination with the JWST pipeline version 1.17.1 and CRDS context version `\texttt{jwst\_1322.pmap}'.
First, the standard pipeline stages were applied with the default settings for stage 1 (\texttt{detector1}), and the following settings for stages 2 and 3.
In stage 2, the image-to-image background subtraction method was applied (`\texttt{bkg\_subtract}'), using the `\texttt{rate}' files obtained by applying stage 1 to the dedicated background observations for this program.
The `\texttt{residual\_fringe}' step was enabled, which reduced the residuals of fringes due to the reflections within the detector \citep{2020A&A...641A.150A}.
In stage 3, we enabled the `outlier\_detection' step, and disabled the `cube\_build' step, so that the final products were the fully processed 2D IFU images (`\texttt{crf}' files). 
The default `\texttt{cube\_build}' step produced cubes whose pixel scale and field of view were different for each of the four MIRI MRS channels.
We decided to  impose a common spatial grid across all wavelengths for the data to be stitched. This choice was motivated by  two main  reasons: \textit{i)} focusing on the exact intersection of fields of view with distinct spatial pixel resolutions would have lead to a significant loss of information; \textit{ii)} because of lower spatial resolution, the data at  longer wavelengths carries less information (in terms of numbers of pixels),  which would have biased the results. This resampling could have been performed after data reduction during the preprocessing step (see below) by a naive spatial interpolation. Instead, this resampling was performed during the data reduction to benefit from the functionalities offered by the pipeline which performs this interpolation more efficiently, using both spatial and spectral information. 
Therefore the `\texttt{cube\_build}' step was performed with custom settings, starting from the stage 3 `crf` files.
The full specification of these settings was as follows. The position angle was set to `\texttt{cube\_pa=250.4}', which aligned with the orientation of the 9x1 mosaic. The pixel scale and field of view were set to those of channel 1 using `\texttt{scale\_xy=0.13}', `\texttt{nspax\_x=33}' and `\texttt{nspax\_y=39}', and the center coordinates were set to `\texttt{ra\_center=83.834782}', `\texttt{dec\_center=-5.418207}'. For channel 1, the resulting cube had nearly identical properties to the one built with the default coordinate system. For the longer wavelength channels, the pixel scale of these custom cubes substantially oversampled the instrumental resolution; this was the expected type of input for the method presented in this work.

\paragraph{Preprocessing.} 

The \textit{Haute Couture} stitching method proposed in Section \ref{sec:method} assumes that the spectral grids between two adjacent cubes are the same for the shared wavelength range, which is not the case for the data provided by the reduction pipeline. As such, the data was spectrally resampled on a common wavelength grid with a consistent step using a linear 1D interpolation.


Additionally, the data contains saturated frames, spikes and so-called bad pixels (i.e., outliers) that needed to be cleaned. Instead of merely discarding the saturated frames, the saturated frames were replaced by corrected frames resulting from a  spectral linear interpolation to avoid gaps with no data for some spectral bands. This step was shown to be of major importance: if the data cubes to be stitched contained one saturated frame, the stitched result might be significantly distorted, e.g., containing unexpected spatial patterns and significant spectral bias. Regarding the remaining spikes and bad pixels in each frame, they were automatically detected and corrected. A significant number of those anomalies exhibited structures with high intensity. To identify them, we implemented a robust anomaly detector following the strategy proposed by \citet{anderson2011}. In each frame $\ve{X}^{c,b}_{i,:}$ associated with a given channel $c$ and sub-channel $b$, the median pixel value $\bar{x}^{c,b}_{i} = \mathrm{median}(\ve{X}^{c,b}_{i,:})$ was first computed as a robust counterpart of the mean pixel values. Then each pixel value significantly far from this robust mean was identified as an anomaly. More precisely, a pixel value ${x}^{c,b}_{ij}$ such that $|{x}^{c,b}_{ij}-\bar{x}^{c,b}_{i}|>t \times \mathrm{MAD}(\ve{X}^{c,b}_{i,:})$ was flagged as an anamoly, where $\mathrm{MAD}(\ve{X}^{c,b}_{i,:}) = \mathrm{median}({x}^{c,b}_{i,j}-\bar{x}^{c,b}_{i})$ computes the median absolute deviation and $t$ is a cutoff parameter that needs to be adjusted for each channel and sub-channel. The pixel values detected as anomalies were finally replaced by the average of their four spatially nearest neighbors. 

Finally, we created a mask containing only the most informative pixels. A pixel value is flagged as not carrying any information if its value is \texttt{NaN}. For each cube, the mask corresponds to the spatial pixels that have more than 99\% spectral information (i.e. that are not \texttt{NaN}).
In the mask, the star at the top of our field of view is removed because it is saturated in most of the wavelength range and the spectra are not exploitable.

\subsection{Choice of the parameters and initialization} \label{sec:choice}

As discussed in Section \ref{sec:method}, \textit{Haute Couture} requires a few hyper-parameters to be set by the end-user. {A first hyper-parameter is the value of $\beta$, i.e., the shape parameter of the divergence used as a measure of fit in NMF. As thoroughly discussed by \citet{fevotte2011}, the $\beta$-divergence is a log-likelihood in disguise, and choosing $\beta$ is similar to making a noise assumption. The range of practical values is generally $\beta \in [0,2]$, with $\beta = 0$ corresponding to multiplicative Gamma noise and $\beta =2$ corresponding to additive Gaussian noise. Choosing $\beta=1$ (generalized Kullback-Leibler divergence), corresponding to a Poisson-like continuous distribution, has been shown to offer an excellent trade-off in many applications. It was in particular advocated in the previous work by \citet{berne2007} for source separation of similar mid-IR data using NMF.}

{A second hyper-parameter to set is $K$, i.e., the rank of the factorization (the common dimension of $\ve{W}$ and $\ve{H}$). A suitable value needs to capture the latent structure in the data while preventing from over-fitting (i.e, modeling the noise rather than the inherent phenomena that explain the data). We followed a trial and error approach by assessing the quality of the reconstruction with a set of candidate values for $K$, namely $K \in [2,4,6,10,15]$. We considered only channel 2 and 3 in order to save computing time. For every couple of estimated matrices $\ve{W}$ and $\ve{H}$} we then calculated the correlation between the reconstruction and the MIRI-MRS spectra (spectra after preprocessing in order to compare) for each sub-channel and the average of the sub-channel. For all $K$, the average correlation was 0.935 $\pm$ 0.0006. The algorithm was not very sensitive to the value of $K$ in our example and, thus, we chose $K=6$ as it maximized the correlation.

{Besides choosing the hyper-parameters $\beta$ and $K$, \textit{Haute Couture} requires choosing an initialization strategy for $\ve{W}$ and $\ve{H}$ in the NMF algorithm. The performance of three methods were empirically assessed: \textit{i)} random initialization, \textit{ii)}  using the outputs of the K-means clustering algorithm, \textit{iii)} ) using the Maximum Angle Source Separation (MASS) technique proposed by \citet{boulais2021}. As for the adjusting the hyper-parameters $\beta$ and $K$, we conducted an exhaustive set of experiments to assess the performance of each of these three methods in reaching the best fit, with $\beta=1$, $K=6$ and again using channel 2 and 3 to save computing time. Our experiments revealed that K-means and MASS return similar results and much better performance than random initialization. As such we favored MASS in the  experiments as it was initially designed for infrared astronomical hyperspectral cubes.}


\begin{figure}[h]
    \centering
    \includegraphics[width=1\linewidth]{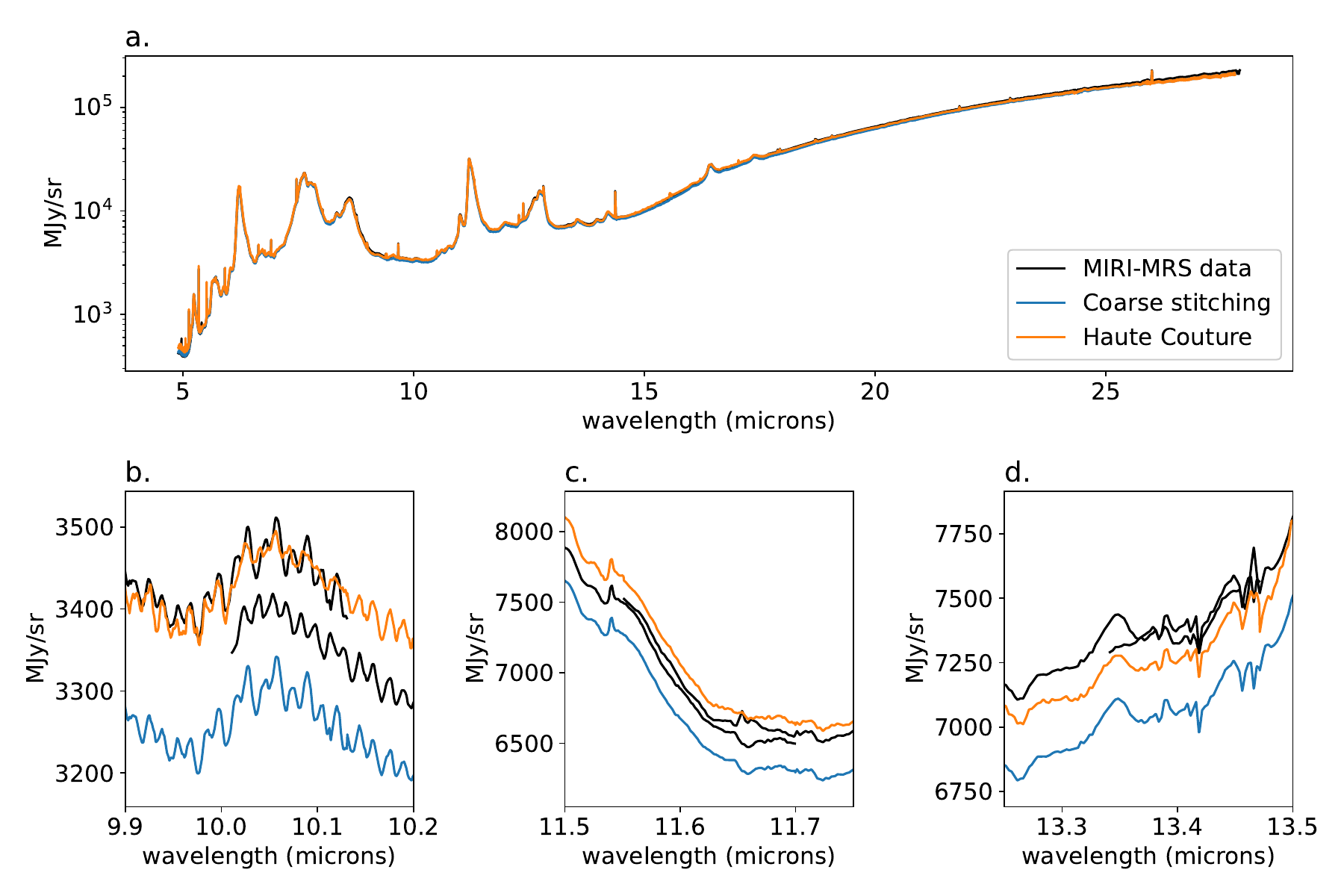 }
    \caption{Top: average spectrum of the MIRI-MRS data after removing the saturated frames, spikes and  bad pixels (black curve), average of the stitched data recovered by \textit{Haute Couture} (orange curve) and average of the result of the coarse stitching procedure (blue curve). Bottom: zooms in spectral ranges where overlap occurs between contiguous sub-channels (as already depicted in Figure \ref{fig:gaps}).}
    \label{fig:spectra}
\end{figure}

\section{Results}\label{sec:results2}

\begin{figure*}[h]
    \centering
    \includegraphics[trim=2 30 100 5,clip,width=0.8\linewidth]{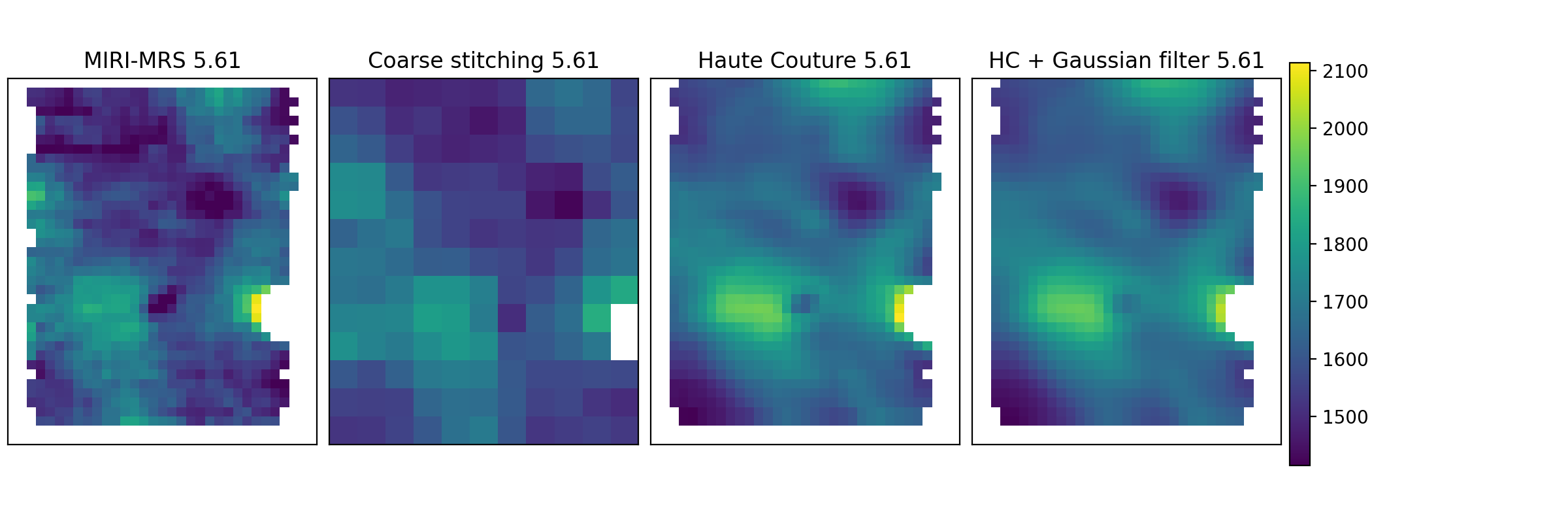}\\
    \includegraphics[trim=2 30 100 5,clip,width=0.8\linewidth]{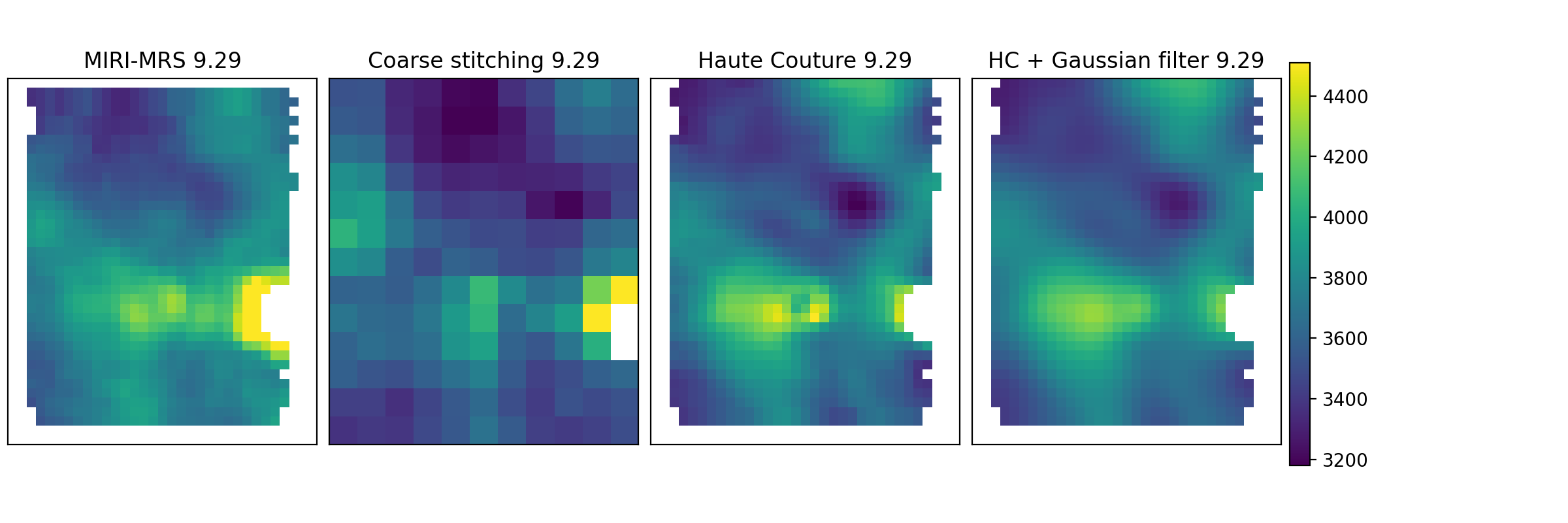}\\
    \includegraphics[trim=2 30 100 5,clip,width=0.8\linewidth]{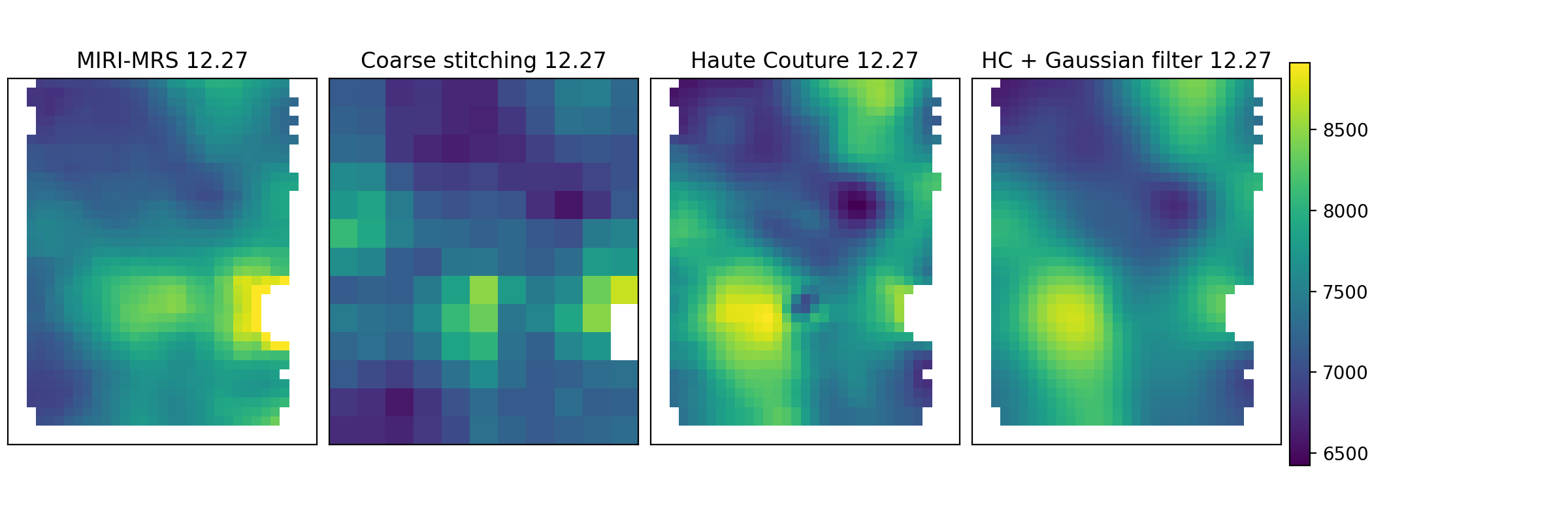}\\
    \includegraphics[trim=2 30 100 5,clip,width=0.8\linewidth]{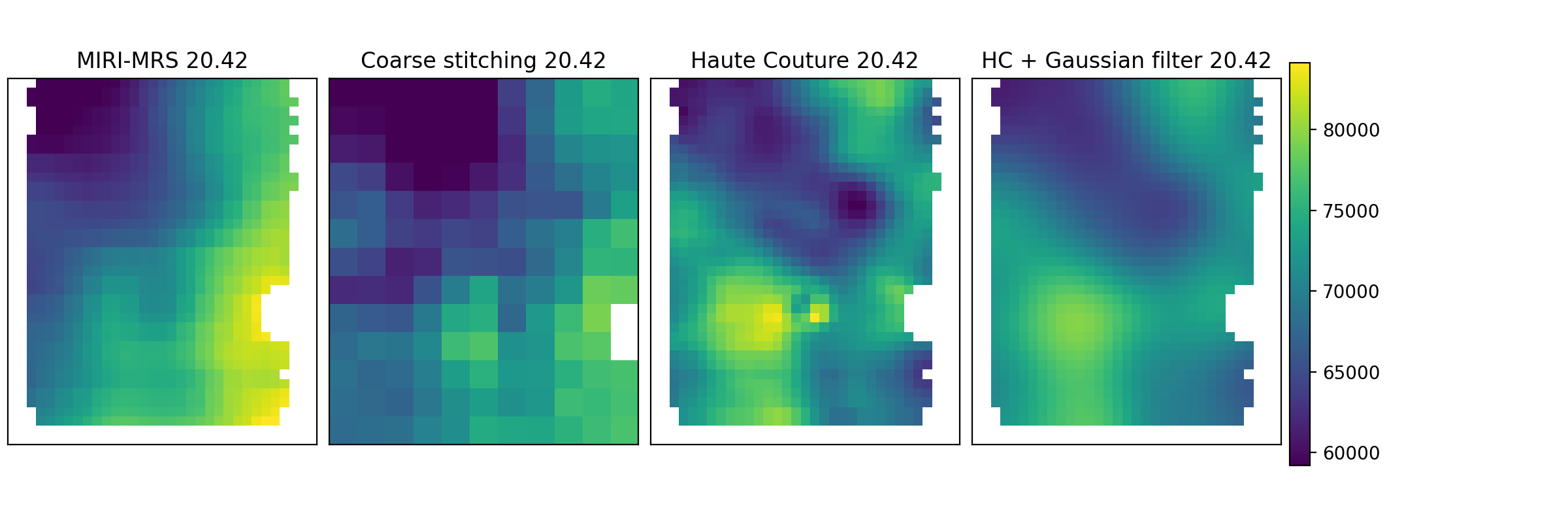}%
    \caption{{Comparison between the MIRI-MRS data ($1$st column), the result of the coarse stitching procedure ($2$nd column), the stitched data produced by \textit{Haute Couture} ($3$rd column) and this final product convolved with a Gaussian filter ($4$th column) in each channel at a given wavelength. The same color bar (in MJy/sr) are used in each row and the data in columns $1$, $2$ and $4$ share the same mask. The loss of spatial resolution is also clearly visible when performing a coarse stitching.}}
    \label{fig:img}
\end{figure*}

\subsection{Stitching results}\label{subsec:results}

{Figure \ref{fig:spectra} (top) displays the average spectrum of the MIRI-MRS data over the entire field of view after removing the saturated frames, the spikes and the bad pixels as explained in Section \ref{sec:data_preprocessing}. It  depicts the average spectrum of the stitched data $\hat{\ve{S}}$ recovered by \textit{Haute Couture} (corresponding to the values in channel 1-{\it long} of the submatrix highlighted in the blue shaded area of Figure \ref{fig:matrice}). This result is compared to the product given by the coarse stitching procedure explained in Appendix~\ref{appx:coarse}. The figure also displays three zooms in the spectral ranges where overlap occur between contiguous sub-channels (bottom). By design, the \textit{Haute Couture} stitched data average spectrum is continuous, i.e., the spectral gaps exhibited in the original MIRI-MRS data have been removed. It is also worth noting that that the fit between the original and \textit{Haute Couture}  reconstructed data is rather tight over the full spectral range.

{Figure \ref{fig:img} displays the preprocessed MIRI-MRS data before stitching and compares the final products resulting from the coarse stitching procedure and from the proposed \textit{Haute Couture} method. This figure also depicts the \textit{Haute Couture} product after undergoing a convolution with a Gaussian filter whose standard deviation $\sigma = \frac{1.22 \lambda}{2\sqrt{2\ln{2}}D}$  has been specifically designed to mimic the effect of the MIRI PSF,  where $\lambda$ is the considered wavelength and $D$ is the mirror diameter. The resulting convolved product can then be also fairly compared to  the original MIRI-MRS data as a validation step.}

{In the first row, corresponding to data and results at $5.61 \mu m$ (channel 1), the \textit{Haute Couture} image appears as significantly denoised and exhibits the same spatial resolution as the original data. This was expected since, as motivated in Section \ref{sec:completion}, the  \textit{Haute Couture} stitching procedure targets the spatial and spectral resolutions of channel 1-{\it long} (blue shaded area in Figure \ref{fig:matrice}). 
{Moreover, the observed denoising effect can be attributed to the low-rank factor model underlying \textit{Haute Couture}. By implicitly representing the spectral data into a $K$-dimensional subspace, the proposed stitching approach preserves the essential physical information while effectively filtering out most of the noise.} 
The second row associated with results at 9.29$\mu m$ (channel 2) shows a lightly denoised and more contrasted \textit{Haute Couture} image compared to MIRI-MRS image. At 12.27$\mu$m (channel 3), the \textit{Haute Couture} image is sharper and more contrasted than the MIRI-MRS data. Interestingly, the spatial structure of the 203-506 protoplanetary disk, known to be present in the considered astrophysical scene \cite{berne2024}, is well reconstructed. The MIRI-MRS image at 20.42$\mu$m (channel 4) does not contain any particular spatial structure and the 203-506 protoplanetary disk is no longer visible. In the \textit{Haute Couture} image, the spatial structures are well recovered.}

{Conversely, in all channels, the product resulting from the coarse stitching procedure is affected by a worse contrast and a clear under-estimation  of the spectral flux density already highlighted in Figure \ref{fig:spectra}. It is also noting that the loss in spatial resolution is significant. This was expected since the coarse stitching procedure targets the poorest resolution offered by the data cubes to be stitched.}

{The most straightforward way to validate an algorithm is to test it against ground truth data, as it is often possible with Earth observation data. However, in the case of astronomical data, obtaining ground truth is impossible. A widely used alternative consists in validating the algorithm using simulated data. This strategy has been followed for instance by \citet{guilloteau2020simulated} to validate data fusion algorithms by generating simulated NIRCam and NIRSpec data. Simulating MIRI-MRS data is, however, more challenging \citep{hadj2017}. The \texttt{Pandeia} tool \citep{pontoppidan2016} can produce simulated MIRI-MRS data, but only allows to simulate simple scenes. The MRISim tool \citep{klaassen2021} allows for more complex scenes, however to our knowledge, the public version of this software has not been updated after the launch of JWST, and therefore the tool may not provide data that is directly comparable to real observations.}
{In this context, as a sanity check, we compare in Figure \ref{fig:img} the results of  \textit{Haute Couture} degraded data after undergoing the Gaussian smoothing (column 4) and those of the original MIRI-MRS data (column 1). It can be seen  there is a good qualitative agreement in terms of the spatial textures between images of column 1 and 4, especially in channels 1 and 2. 
At longer wavelengths, the images differ notably in the right part of the scene. This can be explained by the masking procedure implemented during the preprocessing, which removes pixels associated with the star located in this area (see Section \ref{sec:data_preprocessing}).}

\subsection{Error propagation}

The MIRI-MRS cubes provided by the pipeline as detailed in Section \ref{sec:data_preprocessing} and stitched in Section \ref{subsec:results} are provided with so-called error cubes (specified by the extension \texttt{ERR}). More precisely, each data cube $\ve{X}^{c,b}$ composing the matrix $\ve{X}$ depicted in Figure \ref{fig:matrice} is granted with an error matrix denoted $\Delta \ve{X}^{c,b}$. To quantify the relative amount of error contained in the original MIRI-MRS data to be stitched, we first evaluate the signal-to-error ratio (SER) defined 
as
\begin{align}
    \mathrm{SER} & = 
    10\log \frac{\sum_{c,b} \left\lVert  \ve{X}^{c,b} \right\rVert^2_2}{\sum_{c,b}\left\lVert  \Delta\ve{X}^{c,b} \right\rVert^2_2} = 53.09\textrm{dB}.
\end{align}

\begin{figure}[h!]
    \centering
    {\includegraphics[width=.99\columnwidth]{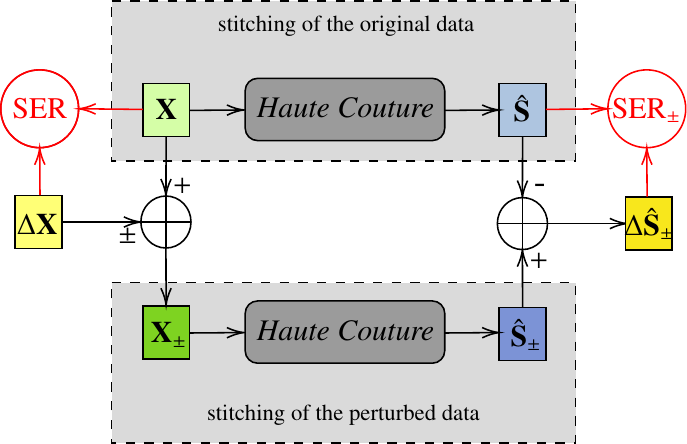}}
        \caption{Quantification of the error propagation. The amount of error $\textrm{SER}$ before stitching is evaluated from the original data contained in $\ve{X}$ (light green box)  and the error matrix $\Delta \ve{X}$  (light yellow box). The original data $\ve{X}$ is processed by \textit{Haute Couture} to produce the stitched spectra $\ve{S}$ (light blue box). The error matrix $\Delta\ve{X}$  is combined with the original data $\ve{X}$ to produce the perturbed data $\ve{X}_{\pm}$ (dark green box). The perturbed data $\ve{X}_{\pm}$ is processed by \textit{Haute Couture} to produce the stitched spectra $\hat{\ve{S}}_{\pm}$ (dark blue box), yielding the propagated error $\Delta \hat{\ve{S}}_{\pm}$ (dark yellow box). The amounts of error $\textrm{SER}_{\pm}$ after stitching is evaluated  from the stitched original spectra  $\hat{\ve{S}}$  and the propagated errors $\Delta \hat{\ve{S}}_{\pm}$.}
    \label{fig:propagation}
\end{figure}

We then propose to assess the propagation of this error when performing stitching. To do so, we introduce the set of data matrices $\ve{X}^{c,b}_+ = \ve{X}^{c,b} + \Delta \ve{X}^{c,b} $ and $\ve{X}^{c,b}_- = \ve{X}^{c,b} - \Delta \ve{X}^{c,b} $ corresponding to the original data positively and negatively perturbed by the error term in each channel $c$ and sub-channel $b$. These submatrices are rearranged to form the incomplete matrices $\ve{X}_+$ and $\ve{X}_-$ as perturbed counterparts of the original incomplete matrix $\ve{X}$ introduced in Section \ref{sec:method} (see also Figure \ref{fig:matrice}). The completion procedure detailed in Section \ref{sec:mcomp} is applied on these two incomplete matrices $\ve{X}_+$ and $\ve{X}_-$, yielding the pair of stitched data $\hat{\ve{S}}_{+}$ and $\hat{\ve{S}}_-$ recovered by the proposed method \textit{Haute Couture}. The propagation of the error is finally evaluated by quantifying how much the original amount of error translates into the perturbed data after undergoing stitching. Formally, we compute the SER associated with the positively and negatively perturbed data as
\begin{equation}
    \mathrm{SER}_+  =  10\log \frac{\left\lVert  \hat{\ve{S}} \right\rVert^2_2}{\left\lVert {\Delta \hat{\ve{S}}}_{+} \right\rVert^2_2} \quad \text{and} \quad     
    \mathrm{SER}_-  =  10\log \frac{\left\lVert  \hat{\ve{S}} \right\rVert^2_2}{\left\lVert {\Delta \hat{\ve{S}}}_{-} \right\rVert^2_2}
\end{equation}
where ${\Delta\hat{\ve{S}}}_{\pm} = \hat{\ve{S}}_{\pm} - \hat{\ve{S}} $ measures the discrepancy between the  stitched spectra $\hat{\ve{S}}$ resulting from the original data  and the stitched spectra $\hat{\ve{S}}_{\pm}$ resulting from the perturbed data, and is considered as the propagated error. We obtain  $\mathrm{SER}_+ = 33.43$dB and $\mathrm{SER}_- = 34.25$dB. The overall procedure is sketched in Figure \ref{fig:propagation}. It appears that the SER deteriorates by about $20$dB, which may be explained by the fact that the error terms $\Delta\ve{X}^{c,b}$ do not likely follow the NMF model underlying the completion procedure detailed in Section \ref{sec:mcomp}. Yet, more importantly, it remains that the amount of additional error induced by \textit{Haute Couture} remains of low magnitude with respect to both the original error and the energy of the stitched data.

\section{Conclusion}

{In this paper we presented a new stitching algorithm, coined as \textit{Haute Couture}, which provides near-optimal assembly of MIRI-MRS spectral cubes. The source code for this method is available through 
[to be inserted upon publication].
\textit{Haute Couture} allows the spatial information to be preserved  throughout the overall spectral range covered by the JWST. Applied on real observations, we show that \textit{Haute Couture} is able to recover spatial structures that are not preserved by a coarse stitching procedure. 
The proposed approach relies on the formulation of the stitching task as a matrix completion problem, subsequently solved by nonnegative matrix factorization. In principle, the versatility of this formulation makes the proposed approach applicable to spectral cubes provided by other spatial instruments, such as NIRSpec also embedded on the JWST, or by other ground-based IFUs.}

\begin{acknowledgements}
This work is partially supported by the Artificial and Natural Intelligence Toulouse Institute (ANITI), funded by the France 2030 program under the grant agreement ANR-23-IACL-0002. This work is funded by the Centre National d’Etudes Spatiales (CNES) through the APR program.
\end{acknowledgements}

%
  \bibliographystyle{aa} 
  \bibliography{ArticleHC} 

\begin{appendix}
    \onecolumn
    \section{Joint scale parameter estimation} \label{appx:opti}

This appendix  details the approach introduced in Section~\ref{sec:scaling} and implemented to mitigate intensity gaps observed in adjacent data cubes provided by MIRI-MRS. To ease the presentation of the method, we introduce the one-to-one mapping between the couple $(c,b) \in \{1,\ldots,4\} \times \{ {\tt s}, {\tt m}, {\tt l} \}$ which indexes a given sub-channel $b$ of a given channel $c$ and the integer $i \in \{1,\ldots,12 \}$ which  indexes the sub-matrix composing $\ve{X}$ according to the particular  order represented in Figure \ref{fig:matrice}. Thus in what follows, the sub-matrix $\ve{X}^{c,b}$ is simply denoted as $\ve{X}^{i}$, leading to the alternate representation of the matrix $\ve{X}$ depicted in Figure \ref{fig:notation}.

    \begin{figure}[!h]
    \centering
    \includegraphics[width=0.35\linewidth]{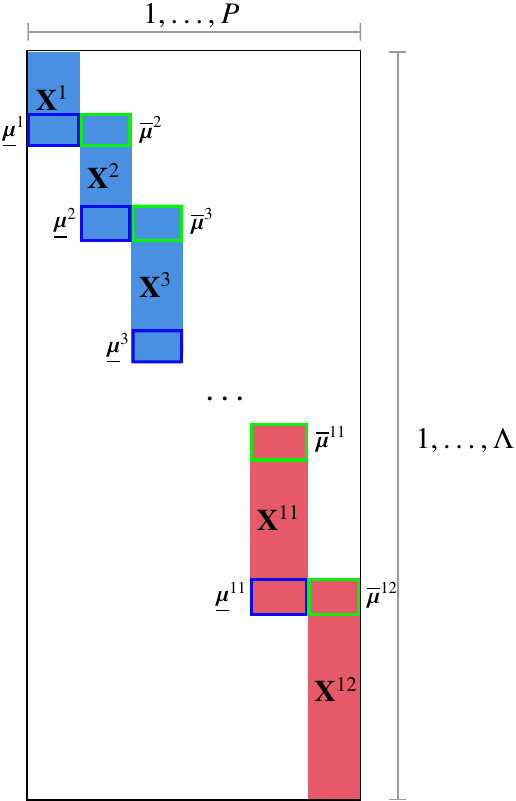}
    \caption{Matrix $\ve{X}$ as of Figure \ref{fig:matrice} with alternative indexing of the sub-matrices.}
    \label{fig:notation}
    \end{figure}

The intensity gap correction aims at estimating a set of twelve scale parameters $\boldsymbol{\alpha} = [\alpha^1,\ldots, \alpha^{12}]^T$ to be applied to the twelve data matrices $\left\{ \ve{X}^1,\ldots,\ve{X}^{12}\right\}$, respectively. By the very specific nature of the data acquisition, the wavelength range associated with the top rows (resp. bottom) of the matrix $\ve{X}^i$ emphasized as green (resp. blue) boxes in Figure \ref{fig:notation} overlaps with the bottom rows of the matrix $\ve{X}^{i-1}$ (resp. top of $\ve{X}^{i+1}$) emphasized as blue (resp. green) boxes in Figure \ref{fig:notation}. It is worth noting that contiguous green and blue boxes frame portions of spectra that correspond to the same spatial locations at the same wavelengths. Thus the proposed approach consists in jointly adjusting the scale parameters by ensuring that the portions of pixel spectra located in the blue boxes match as much as possible the portions of the pixel spectra in the contiguous green boxes. Formally, the spatial average of the top (resp. bottom) rows of $\ve{X}^i$ where such overlap occurs is denoted by $\overline{\bmu}^i$ (resp. $\underline{\bmu}^i$), see Figure \ref{fig:notation}. Maximizing the fit between $\alpha^i \underline{\bmu}^i$ and $\alpha^{i+1} \overline{\bmu}^{i-1}$, for $i=1,\ldots,11$, boils down to solve the optimization problem
\begin{equation}
    \label{eqn:optim}   
    \min_{\boldsymbol{\alpha}}  \sum_{i=1}^{11} \left \lVert \alpha^i \underline{\bmu}^i - \alpha^{i+1} \overline{\bmu}^{i+1} \right \lVert^2_2
\end{equation} 
with $\boldsymbol{\alpha} = [\alpha^1,\ldots, \alpha^{12}]^T$. The problem as formulated by Equation~\eqref{eqn:optim} is ill-posed since one optimal solution is $\boldsymbol{\alpha}^* = [0,\ldots,0]^T$, i.e., setting all the sub-matrices to zero. To remove this trivial, inappropriate solution, one or more sub-channel are served as reference, with scale parameters fixed to an arbitrary value. In particular, in this work we assume that channels 1 and 12 have been properly calibrated and set $\alpha^1 = \alpha^{12} = 1$, but the proposed methodology may apply  with any other arbitrary choices. Thus, we introduce the sub-vector of scale parameters $\tilde{\balpha} = [\alpha^2,\ldots,\alpha^{11}]^T$ and reformulate the problem \eqref{eqn:optim} as
\begin{align} \label{eqn:optim2}   
    \min_{\tilde{\balpha}} \mathcal{J}(\tilde{\balpha}) \quad \text{with} \quad \mathcal{J}(\tilde{\balpha})= \sum_{i=2}^{10} \left \lVert \alpha^i \underline{\bmu}^i - \alpha^{i+1} \overline{\bmu}^{i+1} \right \lVert^2_2.
\end{align}
The criterion $C(\tilde{\balpha})$ to be minimized is smooth and convex: Problem~\eqref{eqn:optim2} has a unique solution that obeys
\begin{equation} \label{eqn:grad}
    \nabla_{\tilde{\balpha}} \mathcal{J}(\tilde{\balpha}) = 0
\end{equation}
where $\nabla$ denotes the gradient operator. Trivial algebra shows that Equation~\eqref{eqn:grad} is equivalent to solving the linear problem
\begin{equation}
\mathbf{M}\tilde{\balpha} = \mathbf{b},
\end{equation}
where $\ve{M}$ and $\ve{b}$ are the square matrix and vector of sizes 10 given by
    \[ \begin{array}{rclcrcl}
        \mathbf{M} & = &  \left ( \begin{array}{ccccc}
            1 & - \frac{\langle \underline{\bmu}^{2},\overline{\bmu}^3 \rangle}{\left \lVert \underline{\bmu}^{2} \right \lVert^2_2 + \left \lVert \overline{\bmu}^{2} \right \lVert^2_2} & 0 & \ldots  & 0 \\
            - \frac{\langle \underline{\bmu}^{2},\overline{\bmu}^3 \rangle}{\left \lVert \underline{\bmu}^{3} \right \lVert^2_2 + \left \lVert \overline{\bmu}^{3} \right \lVert^2_2} & 1 & - \frac{\langle \underline{\bmu}^{3},\overline{\bmu}^4 \rangle}{\left \lVert \underline{\bmu}^{3} \right \lVert^2_2 + \left \lVert \overline{\bmu}^{3} \right \lVert^2_2} & \ddots  & \vdots \\
            0 & - \frac{\langle \underline{\bmu}^{3},\overline{\bmu}^4 \rangle}{\left \lVert \underline{\bmu}^{4} \right \lVert^2_2 + \left \lVert \overline{\bmu}^{4} \right \lVert^2_2} & 1 & \ddots &  0 \\
            \vdots & \ddots & \ddots & \ddots & - \frac{\langle \underline{\bmu}^{10},\overline{\bmu}^{11} \rangle}{\left \lVert \underline{\bmu}^{10} \right \lVert^2_2 + \left \lVert \overline{\bmu}^{10} \right \lVert^2_2} \\
            0 & \ldots & 0 & - \frac{\langle \underline{\bmu}^{10},\overline{\bmu}^{11} \rangle}{\left \lVert \underline{\bmu}^{11} \right \lVert^2_2 + \left \lVert \overline{\bmu}^{11} \right \lVert^2_2} & 1
        \end{array} \right ) &\text{and} &
        \mathbf{b} & = & \left ( \begin{array}{c}
            \frac{\langle \underline{\bmu}^{1},\overline{\bmu}^{2} \rangle}{\left \lVert \underline{\bmu}^{2} \right \lVert^2_2 + \left \lVert \overline{\bmu}^{2} \right \lVert^2_2} \\
            0 \\
            \vdots \\
            0 \\
            \frac{\langle \underline{\bmu}^{11},\overline{\bmu}^{12} \rangle}{\left \lVert \underline{\bmu}^{11} \right \lVert^2_2 + \left \lVert \overline{\bmu}^{11} \right \lVert^2_2}
        \end{array} \right ).
    \end{array}\]
The optimal scale parameters are thus given by $ \tilde{\balpha}^* = \mathbf{M}^{-1}\mathbf{b} $.

\section{Coarse stitching} \label{appx:coarse}

In this section, we present how we produce the coarse stitching presented as comparison in Section \ref{sec:results2}. We use  the MIRI-MRS original data cubes presented in Section \ref{sec:data_preprocessing} and Figure \ref{fig:data}.

\paragraph{Field of view.}The first step is to have a common field of view corresponding to the intersection of each field of view. A exact common spatial grid is impossible due to the pixel size of each channel.

\paragraph{Convolution with PSF.}Each channel and sub-channel is convolved with the PSF of the channel 4-\textit{long} using the library \texttt{stpsf} \citep{perrin2014webbpsf}. We use the Fourier transform to perform the convolution. Then we calibrate the flux in order to have the same flux than before because there is a side effect when convoluting the data that loses flux. This effect is particularly visible at large wavelength because these are less pixels.

\paragraph{Reprojection.} Each channel and sub-channel is reprojected in the spatial grid of the channel 4-\textit{long} using the function \texttt{griddata} of the library \texttt{SciPy}.

\paragraph{Stitching.}The stitching is performed individually for each spatial pixel. The channel 1-\textit{short} is used as reference for the calibration.
For each pixel the average of the fluxes are calculated on the spectral overlap between two successive sub-channel. Then the spectrum that is not the reference is scaled using the ratio of the average.

\end{appendix}

\end{document}